\documentstyle[preprint,aps]{revtex}
\tightenlines
\def\intallk{\int_{-\infty}^{\infty}\frac{dk}{2\pi}}
\def\ef{\left|F\right|}
\def\Poincare{Poincar\'{e} }
\def\F{G\left[\phi\right]}
\makeatother
\begin{document}
\title{An Example of \Poincare Symmetry with a Central Charge}
\author{E. Karat\footnote{E-mail address: karat@mit.edu}}
\medskip
\address{Center for Theoretical Physics,
Laboratory for Nuclear Science
and Department of Physics \\[-1ex]
Massachusetts Institute of Technology, Cambridge, MA ~02139--4307}
\maketitle
\setcounter{page}{1}
\begin{abstract}
\noindent
\baselineskip=16pt

We discuss a simple system which has a central charge in its \Poincare algebra.
We show that this system is exactly solvable after quantization and that
the algebra holds without anomalies.

\end{abstract}
\widetext
\section{Introduction}
Central charges in symmetry algebras arise in two ways.  They can be
anomalies of a quantized theory, like (in two space-time dimensions)
the Virasoro anomaly (triple derivative Schwinger term)
in the diffeomorphism algebra of a diffeomorphism invariant theory
\cite{Cangemi obstruction}\cite{Teitelboim} or in the
(infinite) conformal algebra of a conformally invariant theory
\cite{Fubini}\cite{Gatto}.
However, they can also arise classically, as for example in the
conformally invariant Liouville model, where the (infinite) conformal
algebra possesses a center obtained already by canonical (non-quantal)
Poisson brackets\cite{Poisson brackets} or as in the asymptotic symmetry group
of anti-de Sitter
space in $2+1$ dimensions\cite{anti-de Sitter}.  Another instance arises in the
non-relativistic
field theoretic realization of the Galileo group.

With the appearance of a number of systems with central charges in
their symmetry algebras at the classical level, it is useful to study
a simple model with this behavior.  We examine a charged, scalar field
in 1+1 dimensions, interacting with a constant external electric
field.  The \Poincare algebra of this system has a central charge
appearing already at the classical level.  As a check on \Poincare
invariance, we verify the Dirac-Schwinger relation.  This leads to a
modified energy-momentum tensor.  Next, we take the massless case and
show that we can quantize the system and solve it exactly.  Finally,
we look at the quantized algebra of the massless case and show that it
holds without anomalies, with the electric charge operator functioning
as the central charge.

\section{The Classical Symmetries}

We begin with the Lagrangian in 1+1 dimensions for a complex scalar field of
charge $e$ interacting with a constant external electromagnetic field,
described by a position-dependent vector potential.
\begin{equation}
\label{lagrangian}
L=(D_{\mu}\phi)^*(D^{\mu}\phi)-m^2\phi^*\phi
\end{equation}
with
\begin{eqnarray}
D_{\mu}\phi=\partial_{\mu}\phi+ieA_{\mu}\phi \\
A_{\mu}=-\frac{1}{2}\epsilon_{\mu\nu}Fx^{\nu},
\end{eqnarray}
where we use a flat metric of signature $(+ -)$ and define
$\epsilon^{01}=-\epsilon_{01}=+1$.  Because of the position dependence
in the externally presented vector potential, the theory does not have
manifest translation symmetry.  However, since physical motion is
manifestly translation invariant, we recover invariance of the action
by adding a connection term to the transformation law of the field
$\phi$ under translation.
\begin{eqnarray}
\delta_T^{\mu} \phi&=&(\partial^{\mu}+
\frac{1}{2}ie\epsilon^{\mu\nu}x_{\nu}F)\phi\nonumber \\
&=&(D^{\mu}+ie\epsilon^{\mu\nu}x_{\nu}F)\phi \\
\delta_T^{\mu} L&=&\partial^{\mu}\left[ (D^{\nu}\phi)^*D_{\nu}\phi
-m^2\phi^*\phi\right]
\end{eqnarray}
The associated Noether current $T_C^{\mu\nu}$ is the energy-momentum
tensor for this theory
\begin{eqnarray}
\label{em tensor}
T_C^{\mu\nu}&=&\theta^{\mu\nu}+2A^{\nu}J^{\mu}, \\
\theta^{\mu\nu}&=&(D^{\mu}\phi)^*(D^{\nu}\phi)+(D^{\nu}\phi)^*(D^{\mu}\phi)
-\eta^{\mu\nu}\left((D^{\alpha}\phi)^*(D_{\alpha}\phi)-m^2\phi^*\phi\right), \\
J^{\mu}&=&ie\left[\phi^*(D^{\mu}\phi)-(D^{\mu}\phi)^*\phi\right].
\end{eqnarray}
Using the field equations of motion,
\begin{equation}
\label{eom}
(D^{\mu}D_{\mu}+m^2)\phi=0,
\end{equation}
we see that our currents obey
\begin{eqnarray}
\partial_{\mu}J^{\mu}&=&0 \\
\partial_{\mu}\theta^{\mu\nu}=\partial_{\mu}\theta^{\nu\mu}&=&
\epsilon^{\nu\alpha}FJ_{\alpha} \\
\partial_{\mu}T_C^{\mu\nu}&=&0.
\end{eqnarray}
Note that $T_C^{\mu\nu}$ is not symmetric: it is conserved in the
first index $\mu$, while the second index $\nu$ denotes the direction
of the translation.  Thus we arrive at our first unconventional
result: Even though the fields are spinless, their canonical
energy-momentum tensor is not symmetric.

In addition, the Lagrangian has the usual Lorentz symmetry
\begin{eqnarray}
\delta_L \phi &=& \epsilon^{\alpha \beta}x_{\alpha}\partial_{\beta}\phi
\\
\delta_L L&=&
\partial_{\beta}\left( \epsilon^{\alpha\beta}x_{\alpha}L\right)
\end{eqnarray}
with the associated conserved current
\begin{eqnarray}
\label{lorentz}
M^{\mu}&=&
\epsilon_{\alpha\beta}x^{\alpha}(\theta^{\mu\beta}+A^{\beta}J^{\mu}) \\
\partial_{\mu}M^{\mu}&=&0
\end{eqnarray}

We can define operators on linear functionals $G$ of $\phi$
\begin{eqnarray}
P^{\mu}(\F) &=& G\left[\delta_T^{\mu} \phi \right] \\
M(\F) &=& G\left[\delta_L \phi \right]
\end{eqnarray}
In this way, $P^{\mu}$ generates translations, and $M$ generates Lorentz
transformations.  Defining a bracket to be the commutator of these operators
on linear functionals of $\phi$,
\begin{eqnarray}
\left[M,P^{\mu}\right](\F) &=&
G\left[\delta_T^{\mu}\delta_L \phi - \delta_L\delta_T^{\mu} \phi \right]
\nonumber \\
&=&G\left[\epsilon^{\mu\nu}
(\partial_{\nu}+\frac{1}{2}ie\epsilon_{\nu\alpha}x^{\alpha}F)\phi \right]
\nonumber \\
&=&\epsilon^{\mu\nu}P_{\nu}(\F) \\
\left[P^{\mu},P^{\nu}\right](\F)&=&
G\left[\delta_T^{\nu}\delta_T^{\mu}\phi-\delta_T^{\mu}\delta_T^{\nu}\phi\right]
\nonumber \\
&=&G\left[ie\epsilon^{\mu\nu}F\phi\right] \nonumber \\
&=&-\epsilon^{\mu\nu}FQ(\F),
\end{eqnarray}
where $Q$ is the operator that multiplies by $-ie$, so it simply commutes with
all other operators that act on linear functionals of $\phi$.
\begin{equation}
Q(\F)=G\left[-ie\phi\right]=-ie\F
\end{equation}
Thus, we have a \Poincare algebra in $1+1$ dimensions with a central charge
\cite{group}.
\begin{eqnarray}
\label{alg1}
\left[M,P^{\mu}\right] &=&\epsilon^{\mu\nu}P_{\nu} \\
\label{alg2}
\left[P^{\mu},P^{\nu}\right] &=& -\epsilon^{\mu\nu}FQ
\end{eqnarray}

This algebra can be realized using Poisson brackets with the charges
\begin{eqnarray}
\label{c1}
P^{\mu} &=&\int dx T_C^{0\mu}(t,x) \\
M &=&\int dx M^{0}(t,x) \\
\label{c4}
Q &=&\int dx J^{0}(t,x),
\end{eqnarray}
Since the charges are the spatial integrals of the time components of conserved
currents, the charges are time-independent, assuming that the field $\phi$ dies
off sufficiently rapidly.
(We shall later show that the quantized versions of these operators are
explicitly time-independent in the massless case.)
We calculate $\pi$, the momentum conjugate
to $\phi$, to be $\pi=(D^0 \phi)^*$.  Similarly, the momentum
conjugate to $\phi^*$ is $\pi^*=(D^0 \phi)$.  Writing
$J^{\mu}$, $T_C^{0\mu}$, and $M^{0}$ in terms of these quantities,
\begin{eqnarray}
J^{0}(t,x)&=&ie(\phi^*\pi^* - \phi\pi) \\
J^{1}(t,x)&=&ie((D_1\phi)^*\phi - \phi^*(D_1\phi)) \\
T_C^{00}(t,x)&=& \pi^*\pi + (D_1 \phi)^*(D_1 \phi) + m^2 \phi^*\phi + xFJ^{0}\\
T_C^{01}(t,x)&=& -\pi(D_1 \phi) - (D_1 \phi)^*\pi^* + tFJ^{0} \\
M^{0}(t,x)&=&x\left(\pi^*\pi + (D_1 \phi)^*(D_1 \phi) + m^2 \phi^*\phi \right)
\nonumber \\
&&
+t\left(\pi(D_1 \phi) + (D_1 \phi)^*\pi^* \right)
+\frac{1}{2}(x^2-t^2)FJ^{0}
\end{eqnarray}
we can now calculate the equal-time Poisson brackets we need.
\begin{eqnarray}
\left[J^{0}(x),J^{0}(y)\right]&=&0 \\
\left[J^{0}(x),T_C^{0\mu}(y)\right] &=&
-\epsilon^{\mu\nu}J_{\nu}(y)\delta^{\prime}(x-y) \\
\left[M^{0}(x),J^{0}(y)\right]&=&\left(xJ^{1}(x)-tJ^{0}(x)\right)
\delta^{\prime}(x-y) \\
\left[T_C^{00}(x),T_C^{00}(y)\right]&=&
\left(T_C^{10}(x)+T_C^{10}(y)\right) \delta^{\prime}(x-y) \\
\left[T_C^{00}(x),T_C^{01}(y)\right]&=&
\left(T_C^{11}(x)+T_C^{00}(y)+F(x-y)J^{0}(y)\right)
\delta^{\prime}(x-y)\\
\left[M^{0}(x),T_C^{0\mu}(y)\right]&=&
\left(
-x_{\nu}(T_C^{\nu\mu}(x)+T_C^{\nu\mu}(y))+x^{\mu}{{T_C}^{\nu}}_{\nu}(y)
+\frac{1}{2}(x_{\lambda}x^{\lambda})\epsilon^{\mu\nu}FJ_{\nu}(y)
\right. \nonumber \\ &&\left.
-\epsilon^{\mu\nu}F(x_{\nu}-y_{\nu})x^{\lambda}J_{\lambda}(x)
+y_{\lambda}(x^{\mu}\epsilon^{\nu\lambda}-x^{\nu}\epsilon^{\mu\lambda})
FJ_{\nu}(y)
\right)
\delta^{\prime}(x-y)
\end{eqnarray}
(where the common time argument has been suppressed).
Thus, the charges satisfy
\begin{eqnarray}
\left[Q,M\right]&=&0 \\
\left[Q,P^{\mu}\right]&=&0 \\
\left[M,P^{\mu}\right]&=&\epsilon^{\mu\nu}P_{\nu} \\
\left[P^{\mu},P^{\nu}\right]&=&-\epsilon^{\mu\nu}FQ
\end{eqnarray}
This is the same algebra that we obtained before in (\ref{alg1}-\ref{alg2}).

\section{The Dirac-Schwinger Relation}

The Dirac-Schwinger relation is a method of proving Lorentz invariance.
A system is Lorentz invariant if the energy-momentum tensor
obeys the following condition (for Poisson brackets):
\begin{equation}
\label{dirac-schwinger}
\left[T^{00}(x), T^{00}(y)\right]=\left(T^{01}(x)+T^{01}(y)\right)
\delta^{\prime}(x-y)
\end{equation}
The energy-momentum tensor we obtained in (\ref{em tensor}) obeys
(\ref{dirac-schwinger}) with the indices reversed:
\begin{equation}
\left[T_C^{00}(x), T_C^{00}(y)\right]=\left(T_C^{10}(x)+T_C^{10}(y)\right)
\delta^{\prime}(x-y).
\end{equation}
Unfortunately, $T_C^{01}\neq T_C^{10}$ is not symmetric in its indices, so the
Dirac-Schwinger condition fails.

However, if we modify the energy-momentum tensor to make it
symmetric, the condition then holds.  By adding a
superpotential $\epsilon^{\mu\beta}\partial_{\beta}V^{\nu}$, we obtain
a new energy-momentum tensor
\begin{eqnarray}
T^{\mu\nu} &=& T_C^{\mu\nu} +\epsilon^{\mu\beta}\partial_{\beta}V^{\nu}
\nonumber \\
&=& \theta^{\mu\nu}+\epsilon^{\alpha\nu}x_{\alpha}FJ^{\mu}
+\epsilon^{\mu\beta}\partial_{\beta}V^{\nu}.
\end{eqnarray}
Requiring $T^{\mu\nu}$ to be symmetric, we obtain
\begin{equation}
\partial_{\mu}V^{\mu} = -Fx_{\mu}J^{\mu}.
\end{equation}
Since $J^{\mu}$ is conserved, we can define a new variable $h$ by
\begin{equation}
\label{j}
J^{\mu}=\epsilon^{\mu\nu}\partial_{\nu}h
\end{equation}
With this new expression for $J^{\mu}$, we obtain
\begin{equation}
\label{V}
\partial_{\mu}V^{\mu} = \partial_{\mu}(Fx_{\nu}\epsilon^{\mu\nu}h)
\end{equation}
For solutions to (\ref{V}) and (\ref{j}), we can take
\begin{eqnarray}
V^{\mu} &=& Fx_{\nu}\epsilon^{\mu\nu}h \\
h(t,x) &=& \int_{-\infty}^{\infty}dy \frac{1}{2} \epsilon(x-y) J^{0}(t,y),
\end{eqnarray}
where $\epsilon(x-y)$ denotes the sign function.
For these solutions, $T^{\mu\nu}$ simplifies to
\begin{equation}
\label{symmetric}
T^{\mu\nu} = \theta^{\mu\nu}-\eta^{\mu\nu}hF \\
\end{equation}
Checking our commutation
relations, we see that
\begin{equation}
\left[T^{00}(x), T^{00}(y)\right]=(T^{01}(x)+T^{01}(y))
\delta^{\prime}(x-y),
\end{equation}
so the Dirac-Schwinger condition indeed holds for $T^{\mu\nu}$.

Alternately, we can derive the energy-momentum tensor (\ref{symmetric}) by
varying the metric in
a generally covariant version of the Lagrangian (\ref{lagrangian}).
We start by writing the generally covariant Lagrangian
\begin{equation}
L=\sqrt{-g}(g^{\mu\nu}(D_{\mu}\phi)^*(D_{\nu}\phi)-m^2 \phi^*\phi)
\end{equation}
with $A_{\mu}$ no longer an external quantity, but a functional of the metric
satisfying the relation\cite{Cangemi non-minimal coupling}
\begin{equation}
\label{relation}
\partial_{\mu}A_{\nu}-\partial_{\nu}A_{\mu}=\epsilon_{\mu\nu}F\sqrt{-g}
\end{equation}
Varying the metric, we get
\begin{equation}
\label{variation}
\delta L = \frac{1}{2} \sqrt{-g} \theta_{\mu\nu} \delta g^{\mu\nu}
- \sqrt{-g}J^{\mu}\delta A_{\mu}
\end{equation}
Owing to the covariant conservation of $J^{\mu}$, we can write
\begin{equation}
\label{h}
\sqrt{-g}J^{\mu}=\epsilon^{\mu\nu}\partial_{\nu}h
\end{equation}
as a defining relation for $h$.  Substituting (\ref{h}) in (\ref{variation}) in
the variation of the action and integrating by parts, we get
\begin{equation}
\delta S = \int d^2 x (\frac{1}{2} \sqrt{-g} \theta_{\mu\nu} \delta g^{\mu\nu}
- \epsilon^{\mu\nu} \partial_{\mu}\delta A_{\nu} h).
\end{equation}
Using our relation (\ref{relation}) and simplifying, we end up with our result
\begin{equation}
T_{\mu\nu}=\frac{2}{\sqrt{-g}}\frac{\partial L}{\partial g^{\mu\nu}}
= \theta_{\mu\nu}-g_{\mu\nu}hF
\end{equation}
This lends credence that this is the correct
energy-momentum tensor to use for the Dirac-Schwinger relation.

Unfortunately, the momentum associated with
$T^{\mu\nu}$ differs from the momentum associated with $T_C^{\mu\nu}$.
\begin{equation}
\int dx \left( T^{01} - T_C^{01} \right) = \int dx \partial_1 V^{1} =
\left. V^{1} \right|^{x=\infty}_{x=-\infty}=-tQ
\end{equation}
Furthermore, by taking the time derivative of this difference, we see that
both momenta cannot be conserved simultaneously, except possibly in the
uncharged sector.
(We shall later calculate the charges associated
with our original energy-momentum tensor $T_C^{\mu\nu}$ in the
zero-mass case and see explicitly that they are time-independent.  Therefore,
the momentum associated with $T^{\mu\nu}$ is not conserved, except in the
uncharged sector.)

\section{Solution of the Equations of Motion (for Zero Mass)}

For the remainder of this paper, we shall take our field $\phi$ to be
massless by setting $m=0$.  We shall also absorb $e$ into $F$, replacing
$eF$ by $F$.
Then, we shall use the equations of motion and cannonical commutation relations
to quantize the system.

Putting $\phi$ in the form
\begin{eqnarray}
\label{phi1}
\phi&=&\intallk e^{i(k-\frac{1}{2}Ft)x}
\left(f(T)a_k+f(T)^*b_{-k}^{\dagger}\right) \ef^{-1/4}, \\
\label{T}
T&=&\frac{\epsilon(F)}{|F|^{1/2}}(Ft-k) 
\end{eqnarray}
and setting the mass to zero,
our equations of motion (\ref{eom}) reduce to
\begin{equation}
((\frac{d}{dT})^2+T^2)f(T)=0
\end{equation}
which has a closed form solution in terms of Bessel functions
\begin{eqnarray}
\label{phi2}
f(T)&=&A(k)f_1(T)+B(k)f_2(T), \\
f_1(T)&=&|T|^{1/2}J_{-1/4}(\frac{1}{2}T^2), \\
f_2(T)&=&\epsilon(T)|T|^{1/2}J_{1/4}(\frac{1}{2}T^2).
\end{eqnarray}
$A(k)$ and $B(k)$ are arbitrary functions that parameterize the
creation and annihilation operators, $a_k$ and $b_k^{\dagger}$, which satisfy
the usual commutation relations.

We now impose canonical quatization relations.
Our form for $\phi$ (\ref{phi1}) automatically satisfies
$\left[\phi,\phi^*\right]=\left[\pi,\pi^*\right]=0$.
The condition $\left[\phi,\pi\right]=\left[\phi^*,\pi^*\right]=i\delta(x-y)$
require
\begin{equation}
f(T)f^{\prime}(T)^*-f^{\prime}(T)f(T)^*=i,
\end{equation}
which with the help of the Bessel function identity
\begin{equation}
\label{Bessel identity}
J_{-3/4}(x)J_{-1/4}(x)+J_{3/4}(x)J_{1/4}(x)=\frac{\sqrt{2}}{\pi x}
\end{equation}
implies
\begin{equation}
\label{requirement}
A(k)B^*(k)-B(k)A^*(k)=\frac{\sqrt{2}}{4}\pi i.
\end{equation}
Apart from this constraint
(which can be seen as a normalization condition), the choice of these
functions remains arbitary below.  If one insists on choosing a
parameterization, two useful parameterizations are the constant
parameterization and the parameterization which reduces to the standard
free-field expression in the limit where $F$ goes to zero
\begin{eqnarray}
A(k)&=&\frac{\pi}{4}|F|^{-1/2}|k|
\left(J_{-3/4}(k^2/2|F|)-i\epsilon(Fk)J_{1/4}(k^2/2|F|)\right) \\
B(k)&=&-i\frac{\pi}{4}|F|^{-1/2}|k|
\left(J_{-1/4}(k^2/2|F|)-i\epsilon(Fk)J_{3/4}(k^2/2|F|)\right).
\end{eqnarray}
This second parameterization satisfies the requirement (\ref{requirement})
by the Bessel function identitiy (\ref{Bessel identity}).

\section{Quantization of the Algebra}

In this section, we shall use the original energy-momentum tensor
$T_C^{\mu\nu}$
(\ref{em tensor})
and verify that its associated charges are time-independent.
Inserting the solution for $\phi$ (\ref{phi1}, \ref{phi2}) into the classical
expressions for the charges (\ref{c1}-\ref{c4}),
and normal ordering, we get expressions for the quantized charges
in terms of creation and annihilation operators.
\begin{eqnarray}
Q&=&\intallk e \left[a^{\dagger}_ka_k-b^{\dagger}_{-k}b_{-k}\right] \\
&& \nonumber \\
M&=&\intallk \left[
\left(-k^2/2F-\frac{\sqrt{2}}{\pi}iF
\left(\partial A \partial B^*-\partial B\partial A^*\right)
\right)\left(a^{\dagger}_ka_k-b^{\dagger}_{-k}b_{-k}\right)
\right. \nonumber \\
&&-\frac{1}{\sqrt{2}\pi }iF
\left(A^*\partial B-B\partial A^* + A\partial B^*-B^*\partial A\right)
\nonumber \\
&&{\hskip 0.5 in}
\left(a^{\dagger}_k(\partial a)_k-(\partial b^{\dagger})_{-k}b_{-k}
- (\partial a^{\dagger})_ka_k+b^{\dagger}_{-k}(\partial b)_{-k}\right)
\nonumber \\
&&-\frac{1}{8}F
\left(a^{\dagger}_k(\partial^2a)_k-(\partial^2b^{\dagger})_{-k}b_{-k}
-2(\partial a^{\dagger})_k(\partial a)_k
\right. \nonumber \\
&&\left. {\hskip 1 in}
+2(\partial b^{\dagger})_{-k}(\partial b)_{-k}
+(\partial^2 a^{\dagger})_ka_k-b^{\dagger}_{-k}(\partial^2 b)_{-k}\right)
\nonumber \\
&&+\frac{\sqrt{2}}{\pi}iF\left(\partial A B-A \partial B\right)
\left((\partial a)_kb_{-k} + a_k(\partial b)_{-k}\right)
\nonumber \\
&&\left.+\frac{\sqrt{2}}{\pi}iF\left(\partial A B-A \partial B\right)^*
\left(- a^{\dagger}_k(\partial b^{\dagger})_{-k}
-(\partial a^{\dagger})_kb^{\dagger}_{-k} \right)
\right] \\
&&\nonumber \\
P^{0}&=&\intallk \left[-\frac{\sqrt{2}}{\pi}F
\left(A^*\partial B-B\partial A^* +
A\partial B^*-B^*\partial A\right)
\left(a^{\dagger}_ka_k+b^{\dagger}_{-k}b_{-k}\right)
\right. \nonumber \\
&&+\frac{1}{2}iF
\left(a^{\dagger}_k(\partial a)_k+(\partial b^{\dagger})_{-k}b_{-k}
-(\partial a^{\dagger})_ka_k-b^{\dagger}_{-k}(\partial b)_{-k}\right)
\nonumber \\
&&\left.+\frac{2\sqrt{2}}{\pi}F\left(\partial A B-A \partial B\right)a_kb_{-k}
+\frac{2\sqrt{2}}{\pi}F
\left(\partial A B-A \partial B\right)^*a^{\dagger}_kb^{\dagger}_{-k}\right] \\
&&\nonumber \\
P^{1}&=&\intallk k [a^{\dagger}_ka_k-b^{\dagger}_{-k}b_{-k}]
\end{eqnarray}
We note that these charges are manifestly time-independent,
as claimed earlier.
These charges satisfy the algebra
\begin{eqnarray}
\left[Q,P^{\mu}\right]&=&\left[Q,M\right]=0 \\
\left[M,P^{\mu}\right]&=&-i\epsilon^{\mu\nu}P_{\nu} \\
\left[P^{\mu},P^{\nu}\right]&=&i\epsilon^{\mu\nu}FQ
\end{eqnarray}
with no anomalies.

As a note of caution, we formed the charges first and then calculated
the commutators, rather than calculating the commutators of the local
currents and then integrating over space.  In addition to the Virasoro
anomaly, the currents have other anomalies.  These other anomalies depend
on time as well as the particular parameterization chosen for
$A(k)$ and $B(k)$ in (\ref{phi2}).  Furthermore, when these anomalies
are integrated over space, the results are ill-defined and depend
on how the spatial integration is evaluated.  As an
illustrative example, let us sketch the calculation of the commutator
between the quantized charges $Q$ and $M$ by integrating the
commutator between $:J^{0}:$ and $:M^{0}:$.
\begin{eqnarray}
\left[Q,M\right]&=&\int dx dy \left[:J^{0}(t,x):,:M^{0}(t,y):\right] \\
:M^{0}(t,x):&=&x:\theta^{00}(t,x):-t:\theta^{01}(t,x):
+\frac{1}{2}(x^2-t^2)F:J^{0}(t,x): \\
:J^{0}(t,x):&=&
\int \frac{dk}{2\pi} \frac{dk^{\prime}}{2\pi}e^{i(k-k^{\prime})x}
i\left(\partial_{T_k} - \partial_{T_{k^{\prime}}}\right)
S(k,k^{\prime},T_k,T_{k^{\prime}}) \\
:\theta^{00}(t,x): &=&
\int \frac{dk}{2\pi} \frac{dk^{\prime}}{2\pi}e^{i(k-k^{\prime})x}
\left|F\right|^{1/2}
\left(\partial_{T_k}\partial_{T_{k^{\prime}}}+T_kT_{k^{\prime}}\right)
S(k,k^{\prime},T_k,T_{k^{\prime}}) \\
:\theta^{01}(t,x): &=&
\int \frac{dk}{2\pi} \frac{dk^{\prime}}{2\pi}e^{i(k-k^{\prime})x}
\epsilon(F)\left|F\right|^{1/2}
\left(i\partial_{T_{k^{\prime}}}T_k-i\partial_{T_k}T_{k^{\prime}}\right)
S(k,k^{\prime},T_k,T_{k^{\prime}}) \\
S(k,k^{\prime},T_k,T_{k^{\prime}}) &=&
f_kf_{k^{\prime}}^* a_{k^{\prime}}^{\dagger}a_k
+f_k^*f_{k^{\prime}} b_{-k}^{\dagger}b_{-k^{\prime}}
+f_kf_{k^{\prime}} a_kb_{-k^{\prime}}
+f_k^*f_{k^{\prime}}^* b_{-k}^{\dagger}a_{k^{\prime}}^{\dagger}
\end{eqnarray}
where $T_k$ is the expression (\ref{T}) for $T$, and the added subscript
allows us to denote the same expression with the momentum appearing in the
subscript substituted for $k$.  Similarly, $f_k$ denotes $f(k,T_k)$.
It must also be noted that the independent
variables with respect to the integrals are the momenta ($k$, $k^{\prime}$,
etc.), $x$, and $t$; however, the independent variables with respect to
the derivatives are the momenta, the $T$ variables with respect to each of the
momenta ($T_k$, $T_{k^{\prime}}$, etc.), and $x$.

The operator $S$ satisfies
\begin{eqnarray}
\label{S commutator}
\left[S(k,k^{\prime},T_k,T_{k^{\prime}}),
S(p,p^{\prime},T_p,T_{p^{\prime}})\right] &=&
2\pi\delta(k-p^{\prime})\left(f_kf_{p^{\prime}}^*-f_k^*f_{p^{\prime}}\right)
S(p,k^{\prime},T_p,T_{k^{\prime}}) \nonumber \\
&&-2\pi\delta(p-k^{\prime})\left(f_pf_{k^{\prime}}^*-f_p^*f_{k^{\prime}}\right)
S(k,p^{\prime},T_k,T_{p^{\prime}}) \nonumber \\
&&+2\pi\delta(p-k^{\prime})2\pi\delta(k-p^{\prime})
\left(f_kf_{k^{\prime}}f_p^*f_{p^{\prime}}^*
-f_k^*f_{k^{\prime}}^*f_pf_{p^{\prime}}\right),
\end{eqnarray}
where the delta functions are with respect to the same independent variables
as the integration (momenta, $t$, and $x$).  As a consequence of the different
collections of independent variables, derivatives of the $S$ commutator
(\ref{S commutator}) may not vanish, even though (\ref{S commutator})
vanishes identically.

We can now calculate the other commutators.  The source of the
inconsistency will be most evident if we do not evaluate the integrals over
the delta functions in the anomalous term yet.
\begin{eqnarray}
\left[J^{0}(t,x),J^{0}(t,y)\right]&=&0 \\
\left[J^{0}(t,x),\theta^{00}(t,y)\right]&=&-iJ^{1}(t,y)\delta^{\prime}(x-y)
\nonumber \\
&&+\int \frac{dk}{2\pi} \frac{dk^{\prime}}{2\pi}
\frac{dp}{2\pi} \frac{dp^{\prime}}{2\pi}e^{i(k-k^{\prime})x+i(p-p^{\prime})y}
ie\left(\partial_{T_k} - \partial_{T_{k^{\prime}}}\right)
\nonumber \\
&&{\hskip 0.5 in}
\left|F\right|^{1/2}
\left(\partial_{T_p}\partial_{T_{p^{\prime}}}+T_pT_{p^{\prime}}\right)
\nonumber \\
&&{\hskip 0.5 in}
2\pi\delta(p-k^{\prime})2\pi\delta(k-p^{\prime})
\left(f_kf_{k^{\prime}}f_p^*f_{p^{\prime}}^*
-f_k^*f_{k^{\prime}}^*f_pf_{p^{\prime}}\right)
\\
\left[J^{0}(t,x),\theta^{01}(t,y)\right]&=&-iJ^{0}(t,y)\delta^{\prime}(x-y) \\
\left[J^{0}(t,x),M^{0}(t,y)\right]&=&-i\left(yJ^{1}(t,y)-tJ^{0}(t,y)\right)
\delta^{\prime}(x-y) \nonumber \\
&&+y\int \frac{dk}{2\pi} \frac{dk^{\prime}}{2\pi}
\frac{dp}{2\pi} \frac{dp^{\prime}}{2\pi}e^{i(k-k^{\prime})x+i(p-p^{\prime})y}
ie\left(\partial_{T_k} - \partial_{T_{k^{\prime}}}\right)
\nonumber \\
&&{\hskip 0.5 in}
\left|F\right|^{1/2}
\left(\partial_{T_p}\partial_{T_{p^{\prime}}}+T_pT_{p^{\prime}}\right)
\nonumber \\
&&{\hskip 0.5 in}2\pi\delta(p-k^{\prime})2\pi\delta(k-p^{\prime})
\left(f_kf_{k^{\prime}}f_p^*f_{p^{\prime}}^*
-f_k^*f_{k^{\prime}}^*f_pf_{p^{\prime}}\right)
\\
\left[Q,M\right]&=&
\int dx dy \frac{dk}{2\pi} \frac{dk^{\prime}}{2\pi}
\frac{dp}{2\pi} \frac{dp^{\prime}}{2\pi}ye^{i(k-k^{\prime})x+i(p-p^{\prime})y}
ie\left(\partial_{T_k} - \partial_{T_{k^{\prime}}}\right)
\nonumber \\
&&{\hskip 0.5 in}\left|F\right|^{1/2}
\left(\partial_{T_p}\partial_{T_{p^{\prime}}}+T_pT_{p^{\prime}}\right)
\nonumber \\
&&{\hskip 0.5 in}2\pi\delta(p-k^{\prime})2\pi\delta(k-p^{\prime})
\left(f_kf_{k^{\prime}}f_p^*f_{p^{\prime}}^*
-f_k^*f_{k^{\prime}}^*f_pf_{p^{\prime}}\right)
\\
\label{final form}
&=&\int \frac{dk}{2\pi} \frac{dk^{\prime}}{2\pi}
\frac{dp}{2\pi} \frac{dp^{\prime}}{2\pi}
\nonumber \\
&&{\hskip 0.5 in}
(-i)2\pi\delta(k-k^{\prime})2\pi\delta^{\prime}(p-p^{\prime})
2\pi\delta(p-k^{\prime})2\pi\delta(k-p^{\prime})
\nonumber \\
&&{\hskip 0.5 in}
ie\left(\partial_{T_k} - \partial_{T_{k^{\prime}}}\right)
\left|F\right|^{1/2}
\left(\partial_{T_p}\partial_{T_{p^{\prime}}}+T_pT_{p^{\prime}}\right)
\nonumber \\
&&{\hskip 0.5 in}
\left(f_kf_{k^{\prime}}f_p^*f_{p^{\prime}}^*
-f_k^*f_{k^{\prime}}^*f_pf_{p^{\prime}}\right)
\end{eqnarray}
This final form (\ref{final form}) illustrates the problem with
evaluation -- there are four delta functions of three independent
quantities (four independent momenta, but three independent
differences of momenta).  Evaluating the charges first is equivalent
to evaluating the first two delta functions first.  In this order of
evaluation, the term vanishes.  Evaluating the current commutator
first amounts to evaluating the last two delta functions first.  In
this case, the value of the term depends on the order in which
we evaluate the remaining two delta functions.  This is equivalent to
the value of the term depending on how the $x$ and $y$ integrations
are evaluated.  Since we are interested in the commutators of the
charges, it seems more reasonable to evaluate them completely,
including the spatial integration, before introducing any commutators.
Fortunately, the term whose commutator gives an anomaly has a
vanishing coefficient after the spatial integration.  In this way we
can see that what seems to be an ill-defined anomalous term actually
vanishes.

\section*{Acknowledgements}
I would like to thank Prof. R. Jackiw for suggesting this problem and for
guidance throughout the work.
This work is supported in part by funds provided by the U.S. Department of
Energy (D.O.E.) under cooperative research agreement \#DF-FC02-94ER40818.

\end{document}